\newcommand{\la}{\,\rlap{\raise 0.3ex\hbox{$<$}}{\lower 1.0ex\hbox{$\sim$}}\,}
 \newcommand{\ga}{\,\rlap{\raise 0.3ex\hbox{$>$}}{\lower 1.0ex\hbox{$\sim$}}\,}

\def\@maxsep{}

\documentclass[epsfig,final,mathptm]{aipproc}

\def\selectedlayoutstyle{6x9}
\layoutstyle\selectedlayoutstyle

\SetInternalRegister\hbadness{8000} 

\newcommand\doingARLO[2][]{%
  \ifx\mmref\undefined #1\else #2\fi
}

\def\somespace{\vskip 0.15cm}

\begin{document}

\title 
      [Classical Novae as Super-Eddington Objects]
      {Classical Novae as Super-Eddington Objects}


\author{Nir J. Shaviv}{
  address={Racah Institute of Physics, Hebrew University, Jerusalem 91904, Israel},
  email={shaviv@phys.huji.ac.il}
}

\copyrightyear  {2002}

\begin{abstract}
Several of the inconsistencies plaguing the field of novae are resolved once we consider novae  to be steady state super-Eddington objects. In particular, we show that the super-Eddington shell burning state is a natural consequence of the equations of stellar structure, and that the predicted mass loss in the super-Eddington state agrees with nova observations. We also find that the transition phase of novae can be naturally explained as ``stagnating" winds.
\end{abstract}

\date{\today}

\maketitle

\def\L{{\cal L}}
\def\ledd{{\cal L}_{Edd}}
\newlength{\figurewidth}
\setlength{\figurewidth}{3.7in}
\newlength{\figurewidthsmall}
\setlength{\figurewidthsmall}{2.7in}
\def\mr{\mathrm}
\def\clump{{\cal C}}
\def\func{{\cal W}}
\def\NFH{Nova FH Ser}


\noindent {\bf Introduction:} Classical novae exhibit long duration super-Eddington luminosities while in their eruptive state. At least, this is the conclusion that should be reached when combining that the peak luminosity of novae (with $M_{WD} \ga 0.5 M_{\odot}$, Livio 1992) is always super-Eddington and that in all cases where the bolometric evolution was recovered (using UV observations), it was shown to decay slowly (e.g., Friedjung 1987, Schwarz et al.  2001, Shaviv 2001b). Thus, if classical novae are super-Eddington for durations much longer than any relevant dynamical time scale, two basic questions arise:  

{\parindent 0pt
{\em $\bullet$ How can super-Eddington objects exist for durations much longer than their dynamical time scale?} 

{\em  $\bullet$ Even if such super-Eddington states exist, why do novae choose this state instead of following the known core-mass luminosity relation (CMLR)?}
}

A seemingly unrelated question, which we try to address as well, is 

{\parindent 0pt
{\em  $\bullet$ Why do the theoretical simulations consistently under predict  the amount of ejected mass, as compared with actual observational determinations  of ejecta mass?}}

\vskip 0.2cm
The notion that novae are super-Eddington contradicts the common wisdom usually invoked in which objects cannot shine beyond their classical
Eddington limit, $\ledd$, since no hydrostatic solution exists.  In
other words, if objects do pass $\ledd$, they are highly dynamic. 
They have no steady state, and a huge mass loss should occur since
their atmospheres are then gravitationally unbound and should therefore
be expelled.  Thus, classical novae according to this picture, can
pass $\ledd$ but only for a short duration corresponding to the time
it takes them to dynamically stabilize after the onset of the
thermonuclear runaway (TNR).  This is indeed seen in detailed 1D
numerical simulations of nova TNRs, where novae can be super-Eddington
but only for several thousand seconds (e.g., Starrfield, 1989). 
However, once they do stabilize, they are expected and indeed do reach
 in the simulations a state given by the
CMLR which describes the Hydrogen shell burning state.  Namely, we
naively expect to find no steady state super-Eddington 
atmospheres. This, however, is not the case in nature.


\somespace
\noindent {\bf The Super-Eddington State:} To existence of a super-Eddington state becomes natural, once we consider that:
\begin{enumerate}
\item Atmospheres become unstable as they approach the Eddington limit. In addition to instabilities that operate under various special conditions (e.g., Photon bubbles in strong magnetic fields or the s-mode instability under special opacity laws), two instabilities were found to operate in Thomson scattering atmospheres (Shaviv 2001a). Moreover, one of these instabilities does not depend on the boundary conditions and is therefore extremely general. It implies that {\em all atmospheres will become unstable already before reaching the Eddington limit}.
\item The effective opacity relevant for the calculation of the radiative force on an inhomogeneous atmosphere is not necessarily the microscopic opacity (Shaviv 1998). Instead, it is given by
$    \kappa_{V}^{\mr eff} \equiv {\left\langle F\kappa_{V}\right\rangle_{V}
     / \left\langle F \right\rangle_{V}}  
$.
 The situation is very similar to the Rosseland vs. Force opacity means used in non-gray atmospheres, where the inhomogeneities are in frequency space as opposed to real space. For the special case of Thomson scattering, the effective opacity is always reduced.

\end{enumerate}

To summarize, we find that as atmospheres approach their classical Eddington limit, they will necessarily become inhomogeneous. These inhomogeneities will necessarily reduce their effective opacity such that the effective Eddington limit will not be surpassed even though the luminosity can be super-classical-Eddington. This takes place in the external part of luminous objects, where the radiation diffusion time scale is shorter than the dynamical time scale of the atmosphere. Further inside the object, convection is necessarily excited such that the total energy flux may be super-Eddington, with the radiative part of it necessarily being sub-Eddington and with the convective flux carrying the difference.  

Nevertheless, one of the key features of super-Eddington atmospheres is that a wind will necessarily be accelerated. It is not the catastrophic wind naively expected from super-Eddington conditions, however, it is going to be significant.


\somespace
\noindent {\bf Super-Eddington Winds:} The atmosphere can remain effectively sub-Eddington while being classically super-Eddington, only as long as the inhomogeneities comprising the atmosphere are optically thick. Clearly however,   this assumption should break   at some point where the density is low enough. From that radius upwards, the radiative force overcomes the gravitational pull and a wind is generated.

The mass loss rate can then be obtained by identifying the sonic point of a steady state wind with the critical point, which is the radius where the radiative and gravitational forces balance each other. We then have  $\dot{M} = 4 \pi R^{2} \rho_{critical} v_{sonic}$. Furthermore, this can generally be reduced to the form of  \vskip -0.3cm
\begin{equation}
    {\dot M} = {\func (\Gamma) (L-\L_{\mr Edd}) \over c v_{s}},
    \label{eq:massloss}
\end{equation}
where $\func$ is a dimensionless wind ``function''.  In principle, $\func$
can be calculated from first principles only after the nonlinear state
of the inhomogeneities is fully understood.  This however is still lacking
as it requires elaborate 3D numerical simulations of the nonlinear
steady state.  Nevertheless, it can be done in several 
phenomenological models which only depend on geometrical parameters such as the average size of the inhomogeneities in units of the
scale height ($\beta \equiv d / l_{p}$), the average ratio between the surface area and volume of
the blobs in units of the blob size ($\Xi$), and the volume filling
factor $\alpha$ of the dense blobs.  
For example, in the limit where the blobs
are optically thick, one obtains that
$
    \func = {3 \Xi / 32 \sqrt{\nu} \alpha \beta (1-\alpha)^{2}}
$ (Shaviv 2001b).
Here $\nu$ is the ratio between the effective and adiabatic speeds of sound. 
Thus, $\func$ depends only on 
geometrical factors. It does not depend explicitly on the Eddington 
parameter $\Gamma$.  Moreover, typical values of $\func \sim 1$-$10$ are obtained. 

The mass loss predicted by the super-Eddington theory (eq.~\ref{eq:massloss}) was compared with observations of super-Eddington objects which have good observational data. These were two novae which are not very fast and which have the best determined {\em absolute} bolometric evolution: FH-Ser and LMC 1988 \#1. The theory was also applied to the Luminous Blue Variable star $\eta$-Car. 

For the two novae, we find that the predicted mass loss rates agree with their observations if $\func \approx 10\pm 5$, which is clearly consistent with the theoretical estimate for $\func$. The agreement is also with the temporal evolution of the  velocities,  if those are taken to be the primary absorption line component.  

Using $\func \approx 10\pm 5$, the mass loss equation  can also be applied to $\eta$-Car, which is an entirely different object from novae (in mass, mass loss rate and duration, photospheric size etc), yet, the predicted integrated mass loss is in total agreement with the observed $1$-$2 M_\odot$ of ejecta, while the terminal velocity is consistently predicted as well. For more information on the models, see Shaviv (2001b)


\somespace
\noindent {\bf Steady State Shell Burning:} The next
step is to show {\em why} novae become super-Eddington to begin with.  In particular,
the nova  shell burnings steady state  should
be given by the CMLR (Paczynski 1970), counter to observations.
Given the contradiction,  we look for new solutions to the stellar
structure equations for systems with shell burning.  This should
describe the steady state of novae after they undergo a TNR. Unlike
the standard derivation of the CMLR, we allow the atmospheres to be
inhomogeneous.  Namely, if the Eddington parameter is larger than a
threshold (taken to be 0.85) the effective opacity is reduced.  We still
 do not know the exact behavior of $\kappa_{\mr eff}$
as a function of $\Gamma$.  This could later be obtained by comparing
the steady state obtained to the actual observations of novae, or through detailed 3D radiative hydro simulations.  At
this point however, we want to show that using a reasonable behavior
of $\kappa_{\mr eff}(\Gamma)$, a super-Eddington steady state does
exist.

We take $\kappa_{\mr eff}(\Gamma > \Gamma_{\mr crit}=0.85) = \kappa_{0}
\Gamma_{\mr crit}/\Gamma$ and ${\cal W}=10$ (somewhat different 
choices do not change the conclusions).  We find that the CMLR
obtains a super-Eddington branch.  The main differences between the
structure in this branch and the structure in the sub-Eddington branch
are the following: (1) The super-Eddington branch has a SEW at the top
of it, with the photosphere located in the wind.  The sub-Eddington
branch has no wind (though it could have one if a non-Thomson opacity
is considered).  Since the wind is ``heavy'' the actual luminosity at the
photosphere could be sub-Eddington.  (2) The super-Eddington branch
has a convective layer that penetrates into the burning shell (most of
the energy is actually released in the convection zone).  In the
sub-Eddington branch, the burning shell is all radiative.  (3) The
luminosity in the sub-Eddington branch is not a function of the
core mass.  It is a function of the core mass in the super-Eddington
branch.

Except for the very fast novae, the shell burning evolution can then be described as a steady state which slowly evolves due to mass loss (e.g., as is done by Kato \& Hachisu 1994, for sub-Eddington evolution and opacity driven winds). The preliminary results show that predictions for $L(t)$, $T(t)$, $v_{\infty}$ and $M_{\mathrm ejecta}$ agree with observations.

\begin{figure}[tbh]
 \includegraphics[width=3.6in]{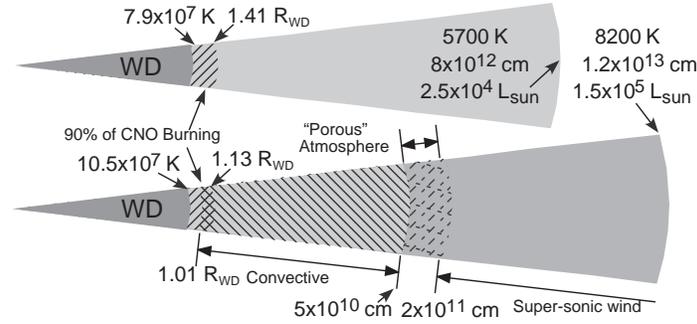}
\caption{
  The shell structure of a CNO shell burning white dwarf with a 
  $10^{-4} M_{\odot}$ envelope. Unlike the top panel, the bottom panel 
  describes an atmosphere in which the radiative instability is taken 
  into account, such that it can become porous. This allows the occurrence 
  of the described steady state. The luminosity is super-Eddington. 
  A convection zone arises and its inner boundary is well within the 
  nuclear burning zone. A continuum driven wind is launched such 
  that the photosphere is in a wind.
  }
  \label{fig:FHTemp}
\end{figure}

%


\somespace
\noindent {\bf A Transition Phase?} An interesting situation can potentially arise if the mass loss predicted requires too large a luminosity to push the ejecta to infinity. This situation is in fact a realization of  the prediction by Owocki and Gayley (1997),  who first suggested that a hypothetically large mass loss rate would result with a stagnating wind. Since no steady state can obviously be obtained, they suggested that a variable state will exit. 

The first analysis performed is a 1D numerical hydro-simulation, which will be elaborated elsewhere (Owocki \& Shaviv  2002). A sample result is given in fig.~2. Although there are no indications that the structure should be purely radial, the limited (and preliminary) 1D results appear to qualitatively  give a typical transition phase behavior. Thus, we hypothesize that this is the origin of the transition phase (excluding of course the big dips which presumably arise from dust formation).

\begin{figure}[tbh]
\includegraphics[angle=-90,width=3.5in]{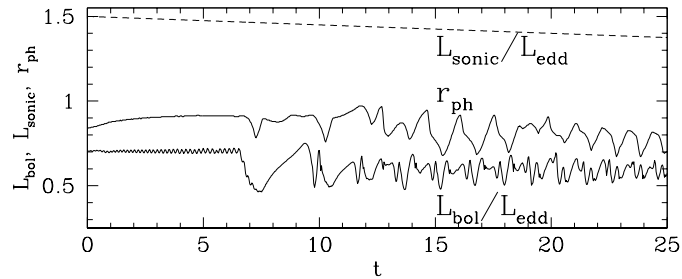} 
\caption{
 Preliminary results of 1D hydro-simulations (Owocki \& Shaviv  2002) where  the luminosity of a system is slowly reduced, such that at some point the wind stagnates before reaching $r\rightarrow\infty$. The radial and temporal units are $10 r_{sonic}$ and $r_{sonic} / v_{s}$. The result resembles the behavior in the transition phase.}
  \label{fig:FHTemp}
\end{figure}


\somespace
\noindent {\bf Discussion:} To summarize, several of the open questions pertaining to classical nova eruptions are resolved if we understand them as steady state super-Eddington objects. As such, the observed super-Eddington luminosities, the temperature and velocity evolution, as well as the mass loss rate are all naturally explained. Moreover, we have seen that by allowing atmospheres to become porous, not only is a steady state possible, it also becomes a viable solution of the shell burning stellar structure equations.

One of the interesting results, is that the velocities predicted from the super-Eddington states are those observed as the primary absorption component. These are typically half as those observed for example in the ``diffuse enhanced'' or ``Orion lines''. Thus, the prediction is that the bulk of the mass outflow which arises in the super-Eddington steady state, should be with the primary velocity. Indeed, if one looks for example at the ejecta of FH Ser years after the eruption, the bulk velocity is similar to the primary component velocity. According to this picture, the higher speed components are therefore significantly less important in the mass loss budget. This however does not solve the question of their origin.


One possibility is that high velocity components are the result of super-Eddington radiative driving of material that is already present outside the critical radius {\em after} steady state super-Eddington shell burning is established, but {\em before} the wind has had time to be established. As a result, this external material is accelerated with the full luminosity, which later falls down with radius as it is ``consumed" to accelerate the primary component (which therefore reaches lower velocities). The pre-existing tenuous material could be a result of a shock wind from the early TNR/dynamic phase.  It is hard to estimate from first principles the mass that should be contained in it, but an upper limit can be given. It cannot be larger than the amount of material that will exist in the steady state wind over a radius of order the sonic radius.

Nevertheless, one can estimate from Shaviv (2000b) the velocity that the high component should have. Optically thin material sitting above the photosphere before the wind begins to accelerate will itself accelerate with the base flux of $\Gamma_0 L_{edd}$. We can compare this to the acceleration in the primary component in which energy was used to up accelerate the ``heavy'' wind, thereby leaving only an energy of flux of $\Gamma_{obs}  L_{edd}$ at $r\rightarrow\infty$. Therefore, at $r\rightarrow\infty$ the initially tenuous material will have a larger terminal kinetic energy such that:
$
{1\over 2} \dot{M} v_\infty^2 \approx {1\over 2} \dot{M} v_{\infty,0}^2 + (\Gamma_0 -\Gamma_{obs})L_{edd}, 
$
where  $v_{\infty,0} $ is the velocity of the steady state wind, that is, of the primary component. Taking the data for FH Ser on day 9 (which is the earliest for which velocity components are present) and {\em nominal} values for the FH Ser super-Eddington state from Shaviv (2000b), we obtain $v_{\infty} \approx 1375$ $km/s$. This is similar to the $1300-1800$ $km/s$ seen for this component.
Moreover, as a wind begins to buildup above the sonic point, initially lower material will be accelerated with a progressively lower flux. This will give rise to a Hubble flow in the tenuous fast component, in which material at larger radii has larger velocities. 

Another noteworthy point is that the integrated ejecta masses observed agree with the mass loss rate predicted by the super-Eddington theory. This implies that if one assumes the envelope masses at TNR to be roughly the observed ejected mass, then consistent nova eruptions are obtained with proper luminosity, temperature, mass loss rate and velocity evolution. The catch is that this envelope mass is typically an order of magnitude larger than the threshold masses required to trigger a TNR. The problem cannot be resolved by ``dredging up" ten times as much mass as the envelope from the WD because the hydrogen fraction seen in the ejecta is not extremely small.
 Thus, we can conclude that the observers properly measure the ejected masses which are consistent with the super-Eddington theory describing the steady state {\em after} TNR. However, we theoreticians do not fully understand the mechanisms regulating the TNR threshold, which is obviously suppressed enough to allow more accumulation of mass before triggering a TNR.

\def\wid{3.45in}
\def\hei{0.65in}
{
\somespace\somespace
\parindent 5pt {\small {\bf TABLE 1:} The emerging "super-Eddington" picture.}}\\ 
{\footnotesize 
\begin{tabular}{ll}
\hline
Phase & Schematic Diagram (not to scale!) \\
\hline
\raisebox{0.1cm}{\begin{minipage}[b]{\wid}
{\bf TNR}. The nova eruption begins with a TNR. Initially, the system is dynamic. Some mass loss can occur from shocks or radiation driving. However, the amount of mass lost this way is insignificant compared to the total mass ejected.
\end{minipage}} & \raisebox{0.0cm}{ \includegraphics[height=\hei]{step1.epsf}} \\
\hline
\begin{minipage}[b]{\wid}
{\bf Formation of the super-Eddington steady state}. Even before the wind reaches an equilibrium, a steady state is reached with shell burning, convection and a ``porous'' envelope. Before the wind builds up, any tenuous material {\em already} present above the critical point will accelerate to {\em high} velocities. 
\end{minipage} &  \raisebox{0.05cm}{\includegraphics[height=\hei]{step2.epsf}} \\
\hline
\begin{minipage}[b]{\wid}
{\bf Acceleration of a thick wind}. Once the critical point is established in the envelope, a super-Eddington thick wind is accelerated. With time, the steady state extends to include the wind. Since the wind is heavy, the radiation field is ``used up''  such that the wind reaches slower (``primary'') terminal velocities. 
\end{minipage} & \raisebox{0.05cm}{ \includegraphics[height=\hei]{step3.epsf}} \\
\hline
\begin{minipage}[b]{\wid}
{\bf Stagnating wind conditions (Optional!)}. If the sonic radius shrinks fast enough, but the luminosity does not, conditions may form for a stagnating wind. That is, a wind too heavy to be pushed to $r\rightarrow\infty$ is predicted such that there is no steady state. This could explain of the  ``oscillatory'' type transition phase.
\end{minipage} &  \raisebox{0.2cm}{\includegraphics[height=\hei]{step4.epsf}} \\
\hline
\begin{minipage}[b]{\wid}
{\bf Sub-Eddington state}. Once a sub-Eddington state is re-established, the thick wind stops and the bare object emerges. The ejected gas is predominantly expanding with the ``primary" velocity. A tenuous part has velocities typically twice as high, with a high dispersion -- its outer parts accelerate to higher velocities. Thus, a Hubble flow will exist in the tenuous material.  
\end{minipage} &  \raisebox{0.25cm}{\includegraphics[height=\hei]{step5.epsf}} \\
\hline
\end{tabular}
}


\vskip -1.0cm 
\begin{thebibliography}{8}

\def\aap{Astron. Astrophys.}
\def\mnras{Mon. Not. Roy. Astron. Soc.}
\def\apj{Astrophys. J.}
\def\apjl{Astrophys. J.}
 
 
    \bibitem{Friedjung1987} {Friedjung}, M. 1987, {\aap}, {179}, 164
    

\bibitem{Kato1994} Kato, M., \& Hachisu, I. 1994, {\apj}, 437, 802    

    \bibitem{Livio1992} {Livio}, M. 1992,
 {\apj}, {393}, 516

\bibitem{Owocki1997} Owocki, S. P., \& Gayley, K. G. 1997, in
 Nota, A., \& Lamers, H., eds, ASP Conf. Ser. Vol. 120: 
Massive Stars in Transition, p.~121

\bibitem{Shaviv2002}  Owocki, S. P., \& Shaviv, N. J., in preparation
 
\bibitem{Paczynski1970} {Paczynski}, B. 1970, {Acta Astronomica}, {20}, 47 
    
   \bibitem{Schwarz2001} {Schwarz}, G.\ J., {Shore}, S.\ N., {Starrfield}, S., {Hauschildt},
 P.\ H., {Della Valle}, M., \& {Baron}, E.\ 2001, {\mnras}, {320}, 103
 
  \bibitem{Shaviv1998} {Shaviv}, N.~J.
 1998, {\apjl}, {494}, L193

 \bibitem{Shaviv2000} {Shaviv}, N.~J.
 2000, {\apjl}, {532}, L137

   \bibitem{Shaviv2001a}  Shaviv, N.~J. 2001a, {\apj}, {549}, 1093
    
   \bibitem{Shaviv2001b} Shaviv, N. J. 2001b, {\mnras}, to appear
    
 \bibitem{Starrfield1989} {Starrfield}, S.
 1989, in {Bode} M.~F., \& {Evans} A., eds, {Classical Novae}. John
 Wiley \& Sons, Oxford
    
\end{thebibliography}
\end{document}